\documentclass[iop]{emulateapj}

\usepackage{amssymb}                    

\shorttitle{Short-term evolution of coronal hole boundaries}
\shortauthors{Krista et al.}

\begin{document}


\title{Short-term evolution of coronal hole boundaries}


\author{Larisza D. Krista, Peter T. Gallagher and D. Shaun Bloomfield}
\affil{Astrophysics Research Group, School of Physics, Trinity College Dublin, Dublin 2, Ireland}



\begin{abstract}

The interaction of open and closed field lines at coronal hole boundaries is widely accepted to be due to interchange magnetic reconnection. To date, it is unclear how the boundaries vary on short timescales and at what velocity this occurs. Here, we describe an automated boundary tracking method used to determine coronal hole boundary displacements on short timescales. The boundary displacements were found to be isotropic and to have typical expansion/contraction speeds of $\leq$2~km~s$^{-1}$, which indicate magnetic reconnection rates of $\leq$3~$\times$~10$^{-3}$. The observed displacements were used in conjunction with the interchange reconnection model to derive typical diffusion coefficients of $\leq$3~$\times$~10$^{13}$~cm$^{2}$~s$^{-1}$. These results are consistent with an interchange reconnection process in the low corona driven by the random granular motion of open and closed fields in the photosphere.

\end{abstract}


\keywords{Sun: corona --- Sun: evolution --- Magnetic fields}



\section{Introduction}

Coronal holes (CHs) are regions of the solar atmosphere with reduced emission and density and hence appear as dark areas at X-ray and extreme ultraviolet wavelengths \citep{Altschuler72, Vaiana76, Chapman02}. As regions of predominantly open magnetic field, CHs are known to be dominated by a single polarity. This characteristic is often used to distinguish CHs from other low-intensity regions \citep{Harvey02, Scholl08, Krista09}. Here, we study the small-scale changes in the CH boundaries that are thought to result from magnetic reconnection occurring between CH open fields and small closed fields found in the quiet Sun (QS) as well as in CHs \citep{Madjarska04}. We investigate the driving force behind the open and closed field interaction and determine the magnetic reconnection rate and the diffusion coefficient based on the observed displacement in CH boundaries.

A number of authors have previously studied the large-scale and long-term evolution of CHs and established the changes in the total CH area, flux, the latitudinal distribution over the solar cycle and the key role CHs play in the dipole-switch of the Sun \citep{Bravo97, Chapman02, Harvey02, Wang96, Wang09}. Although the small-scale changes in CH boundaries have only been studied in detail in recent years using high-resolution observations \citep{Madjarska09, Subramanian10}, it was already suggested in the late 1970s that the small-scale and short-term evolution of CH boundaries is strongly connected to bright point occurrences \citep{Nolte78, Davis85} and that reconnection could be a means of flux-transportation and ``diffusion'' of the open flux \citep{Nolte78}. A more in-depth discussion on reconnection occurring at CH boundaries was provided by \cite{Mullan82}, who proposed that reconnection was also a driver of the solar wind emanating from CHs. 

The process of continuous reconnection at CH boundaries later became better known as ``interchange reconnection'' \citep{Wang93, Wang96, Wang04, Fisk05, Raju05, Edmondson10}. As argued by \cite{Fisk01} and \cite{Fisk05}, the transfer of magnetic flux along the solar surface can be considered as a diffusion-convection process. Despite all of the processes involved being convective in nature, the non-stationary random supergranular convection is included in the diffusion term while those that are uniform (such as differential rotation and meridional flow) are included in the convection term. As supergranular motions are too slow for flux transportation, interchange reconnection was suggested as a faster mechanism for field-line diffusion. Interchange reconnection allows for the transfer of magnetic flux from one point to another through magnetic reconnection between open and closed field lines, which \cite{Fisk01} modelled using a diffusion-convection equation,
\begin{equation}
\frac{\partial B_{z}}{\partial t} = \nabla^{2} (\kappa B_{z}) - \nabla \cdot (uB_{z}) \ ,
\label{eq_fisk}
\end{equation}
where $B_{z}$ is the radial magnetic field, $u$ is the uniform convective flow velocity and 
\begin{equation}
\kappa=\frac {(\delta r)^{2}}{2 \delta t}
\label{kappa}
\end{equation}
is the diffusion coefficient that depends solely on the open field displacement distance ($\delta r$) and the time elapsed ($\delta t$). A graphical representation of interchange reconnection is given in Figure 1, where a QS loop borders the CH open magnetic field. Magnetic reconnection between the open and closed field leads to the displacement of the open field and thus the CH boundary moves further out. During this process the larger loop is replaced by a smaller loop with oppositely oriented footpoint polarities imbedded inside the CH, while the open field line is ``relocated" to the former foot-point of the larger loop. 

\begin{figure}[!ht]
\centerline{\hspace*{0.015\textwidth} 
\includegraphics[angle=-90,scale=0.32, trim=120 40 80 40, clip]{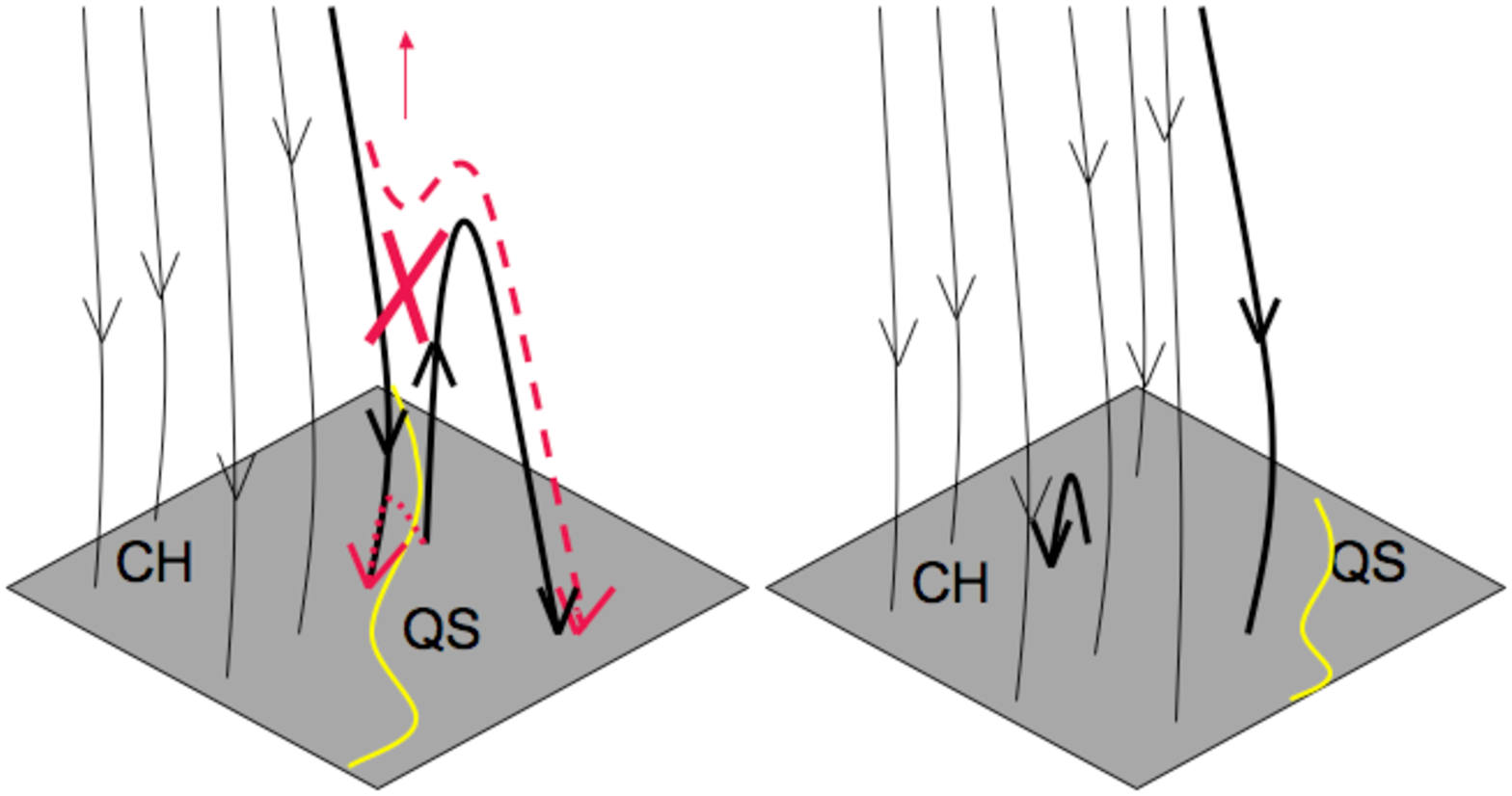} } 
\vspace{-0.01\textwidth} 
\caption{Interchange reconnection: The location of the CH and the QS are indicated on the cartoon with a line representing the boundary between the two regions. The scenario on the left shows a QS loop located near a CH open field which yields magnetic reconnection due to the proximity of the opposite polarity loop footpoint to the open field. The result of the reconnection is shown on the right - a small loop is created inside the CH with opposite magnetic orientation and the open field line ``relocated" from within the CH to the further foot-point of the previously larger loop. This process can explain the rigid rotation as well as the growth of CHs \citep[modified from][]{Fisk05}.}
\label{fig1}
\end{figure} 

Interchange reconnection at CH boundaries is also supported by observational results, e.g., the prevalence of very small loops inside CHs with heights $<$10~Mm and larger loops with heights 10--20~Mm outside CHs \citep{Wiegelmann04}, the presence of bidirectional jets \citep{Madjarska04, Doyle06, Kamio09}, and the boundary displacements observed due to the emergence and disappearance of bright small-scale loops in the form of bright points \citep{Madjarska09, Subramanian10}. While the observed small-scale dynamic processes appear to be in agreement with the interchange reconnection model, it was argued by \cite{Antiochos07} that intermixed open and closed fields of the interchange reconnection model would lead to the formation of currents sheets, which cannot be maintained in a low plasma-beta corona. Instead, the quasi-steady models (e.g. potential field source surface and magnetohydrodynamic models) explain the long-term evolution of the global open magnetic flux by the diffusion of active region magnetic flux and its transport to high latitudes by the meridional flow \citep{Wang00, Wang10}. Also, the quasi-steady models suggest that open field regions of the same polarity are connected by ``corridors", since disconnected open fields (such as those created in the interchange reconnection process) are not allowed \citep{Antiochos07, Edmondson09}. \cite{Antiochos07} suggested that the ``corridors" connecting the open field regions could be wide containing highly dynamic, intermixed open and closed fields which could accommodate interchange reconnection.

In this Letter, we study the short-term evolution of CH boundaries by objectively determining CH boundaries based on intensity thresholding methods, as well as introducing a new automated boundary tracking tool that allows us to follow the changes that occur (detailed in Section \ref{sec:methods}). In Section \ref{sec:results} the magnetic reconnection rate and diffusion coefficient are determined using the interchange reconnection model, and the measured boundary displacement velocities. The processes that influence the CH boundary evolution are discussed in Section \ref{sec:conclusions}.

\section{Observations and Data Analysis}
\label{sec:methods}
\begin{figure}[!ht]
\centerline{\hspace*{0.015\textwidth} 
\includegraphics[angle=90, scale=1.469, trim=-5 0 5 0, clip]{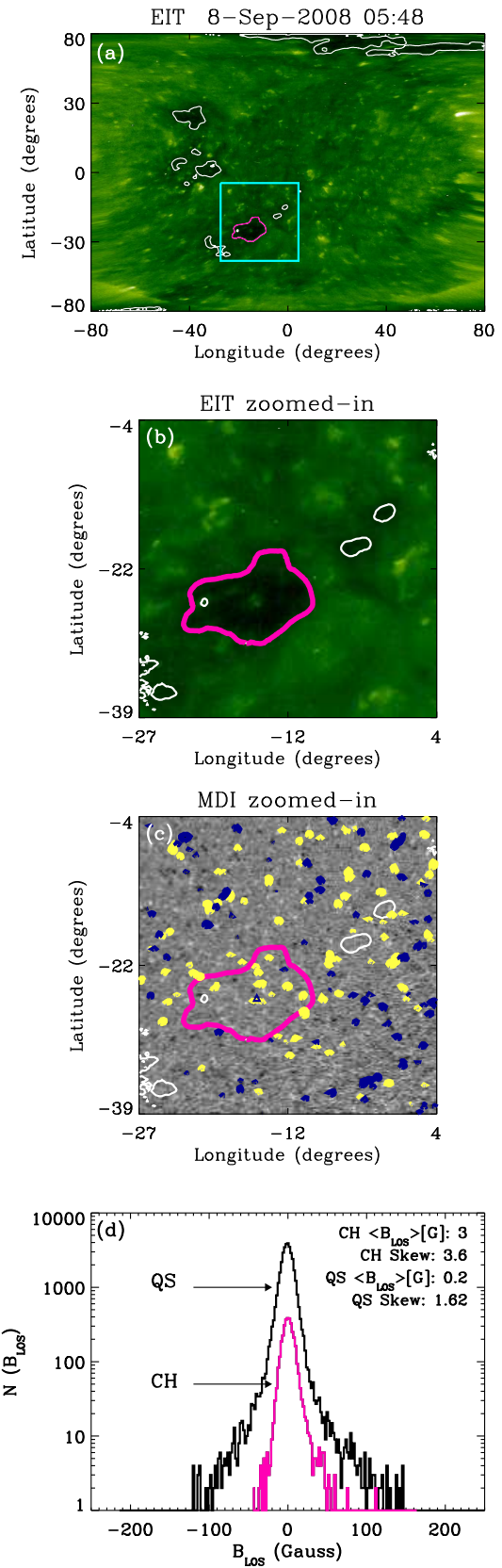} }
\vspace{-0.01\textwidth} 
\caption{CH contours in an EIT 195~\AA\  Lambert projection map (a) and the corresponding sub-image (b). MDI sub-image with the CH contours (c) ($>$50~G fields: yellow, $<$-50~G fields: blue) and the corresponding $B_{\mathrm{LOS}}$ histogram (d) (magenta: CH, black: QS).}
\label{fig2}
\end{figure}

Finding objective CH boundary intensity thresholds is fundamental in deriving consistent physical properties (e.g., area, magnetic field strength) throughout the solar cycle. Many authors have previously chosen arbitrary intensity thresholds to define CH boundaries \cite[e.g.,][]{Baker07, Abramenko09, Madjarska09}. In order to improve the localisation of CH boundaries, \cite{Krista09} developed an automated intensity thresholding algorithm to determine the boundary threshold between CHs and the QS based on local intensity histograms. The same method is used in this study for a smaller region of interest, a 31$^{\circ}$~$\times$~35$^{\circ}$ box centered on each CH studied. Here we use 12~minute cadence 195~\AA\ images from the {\it SOHO} Extreme ultraviolet Imaging Telescope (EIT; \citeauthor{Delaboudiniere95} \citeyear{Delaboudiniere95}) to locate low intensity regions, and 96~minute cadence magnetograms from the Michelson Doppler Imager (MDI; \citeauthor{Scherrer95} \citeyear{Scherrer95}) to detect predominant polarities and hence identify CHs. The magnetograms were also used to determine the magnetic properties of the CHs and the surrounding QS. For this study, eight relatively small CHs were chosen arbitrarily between 2008 September 8 and October 17, and tracked for two to six days. The CHs studied included both predominantly positive and negative polarity holes, with equal numbers located in the northern and southern hemispheres. All of the CHs studied were non-fragmenting and of relatively simple morphology throughout their evolution.
 
The EIT images were initially transformed into Lambert equal-area projection maps to aid the thresholding algorithm and to simplify the tracking of the region of interest. Figure~\ref{fig2}a shows the EIT Lambert projection map with the CH contours and the box around one of the CHs studied (only the largest CH in the box was a candidate in our study). Figures~\ref{fig2}b and \ref{fig2}c show the EIT and MDI close-up images of the box with the CH contours, while Figure~\ref{fig2}d shows the histogram of the line-of-sight magnetic field ($B_{\mathrm{LOS}}$) from the largest CH (magenta line) and the surrounding QS (black line) excluding small satellite CHs. The skewness of the $B_{\mathrm{LOS}}$ histogram is used to establish whether a predominant polarity is found, thus allowing the differentiation of CHs (skewness $>$~0.5) from other low intensity regions (skewness $<$~0.5). The agreement between the visual and the analytically-obtained CH boundary, together with the predominant polarity found in the contoured regions, validate the appropriateness of the boundary positions and justifies the CH status of the detected regions. In principle, the intensity of a CH boundary can change due to the line-of-sight obscuration at higher longitudes and latitudes. However, all CH candidates in this study lie within $\pm50^{\circ}$ in longitude and latitude, which lowers both the obscuration effect in EUV images and the errors in line-of-sight magnetic field measurements.

\begin{figure}[!htb]
\centerline{\hspace*{0.015\textwidth} 
\includegraphics[angle=90, scale=0.8, trim= 15 0 -20 0]{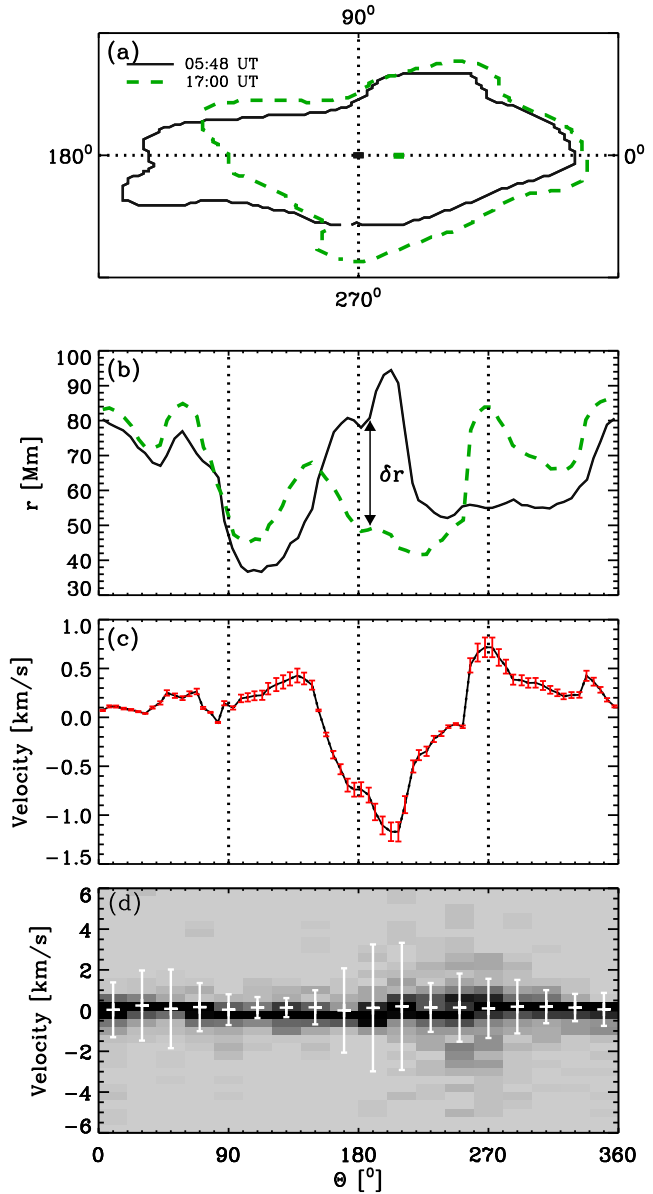} }
\vspace{-0.01\textwidth} 
\caption{(a) The CH contour observed at 05:48 UT (solid line) and 17:00 UT (dashed line) on 2008 September 8. The frame of reference is the earlier contour centroid and angle is measured anti-clockwise from the right direction of the earlier contour centroid. (b) The radial location of the boundary plotted as a function of position angle (solid: earlier; dashed: later). (c) Boundary displacement velocities as a function of position angle. (d) Boundary displacement velocity distributions as a function of position angle for all observations of the CH. The dotted lines in (a), (b), and (c) mark the boundary locations at $0^{\circ}$, $90^{\circ}$, $180^{\circ}$ and $270^{\circ}$. The error bars in (d) correspond to $\pm 1~\sigma$ for each $20^{\circ}$ bin.}
\label{fig3}
\end{figure}

Once a CH was identified, consecutive boundary observations were compared. Considering an example of contour pairs as shown in Figure~\ref{fig3}a, the centroid of the earlier observed contour (solid line, observed on 2008 September 8 at 05:48 UT) was considered as the origin of the frame of reference for both contours. The later contour (dashed line, observed on 2008 September 8 at 17:00 UT) will have moved with the synodic solar rotation rate compared to the earlier contour, due to solar rotation. For this reason a correction was applied to the second contour coordinates. The locations of both contours were measured radially from the earlier centroid in 5$^{\circ}$ segments and anti-clockwise from west of the earlier contour's centroid (Figure~\ref{fig3}b). Each boundary segment location was determined as the median of the contour point distances from the reference centroid. The displacement of a segment ($\delta r$) and the time difference between the two observations yielded the displacement velocity of the segment. This calculation was repeated to get all the boundary displacement velocities over 360$^{\circ}$ (Figure~\ref{fig3}c). The displacement velocities measured for the whole observation period of a single CH were also determined, as shown in the 2D histogram of Figure~\ref{fig3}d. Here, dark regions correspond to more frequent occurrences of the boundary velocity values, where the observations were binned in 20$^{\circ}$ and 0.4~km~s$^{-1}$ intervals. These calculations were further extended to include every boundary displacement velocity of each CH in the study, as detailed in the following Section.

\section{Results}
\label{sec:results}

\subsection{Coronal hole boundary displacement velocities}

In the case of the aforementioned example CH, similar average velocities were observed over all angle space (Figure~\ref{fig3}d). This indicates that there is no preferred direction in the boundary evolution of the CH, and hence the boundary evolution is isotropic. The value of the average signed velocities, $\langle v \rangle$, is $\sim$0~km~s$^{-1}$ in all angle bins, which indicates that there is no consistent expansion or contraction in the CH during the time of observation. Furthermore, the majority of $\pm 1~\sigma$ values for all angle bins are below 2~km~s$^{-1}$ (as is the case for all eight CHs studied here), which is in good agreement with the average velocities of random granular motions. 

In order to extract characteristic velocities, we combine all CH observations over all angles and consider the unsigned boundary velocity distribution (Figure~\ref{fig4}a). From this distribution we determine typical unsigned velocities to be $\leq$2~km~s$^{-1}$. The positive and negative velocities tend to cancel when averaged over time, giving an average signed velocity of $\sim$0~km~s$^{-1}$. However, this does not imply that the CHs studied showed no variation in size during the observing period, rather that the expansion and contraction were on comparable scales.

\subsection{Magnetic reconnection rates}

Magnetic reconnection between CH open fields and small-scale QS loops are expected to occur in the transition region and low corona. Here we can make the assumption that as a small-scale loop reconnects with the open magnetic field in the transition region it will consequently cause a displacement between open and closed field lines in the inner corona. In fact, the displacement at the transition region should be roughly proportional to the displacement detected in EUV images if the surrounding QS has a relatively stable magnetic configuration. The Alfv\'en speed, $v_{\mathrm{A}}$, was calculated using $v_{\mathrm{A}}=B/\sqrt{4\pi m_{p}n_{e}}$, where $B$ is the magnetic field strength and $n_{e}$ is the electron number density. Typical values for the magnetic field ($B=2$~G) and density ($n_{e}=10^{8}$~cm$^{-3}$) in the lower corona were used from \cite{Schrijver05} to calculate the Alfv\'en speed, which was found to be $\sim$600~km~s$^{-1}$. The magnetic reconnection rate was determined from,
\begin{equation}
M_{0}=v_{\mathrm{in}}/v_{\mathrm{A}},
\label{asch}
\end{equation}
where $v_{\mathrm{in}}$ is the inflow speed, the relative speed of the opposite polarity magnetic field lines being brought together prior to reconnection. The distribution of displacement velocities was used for $v_{\mathrm{in}}$ and substituted together with the Alfv\'en speed into Equation~\ref{asch} to obtain the reconnection rate distribution. This distribution is shown in Figure~\ref{fig4}b, where typical reconnection rates are $\leq$3~$\times$~10$^{-3}$.

\subsection{Magnetic diffusion coefficients}
\begin{figure}[!htb]
\centerline{\hspace*{0.015\textwidth}  
\includegraphics[angle=90, scale=0.7, trim= -10 20 0 0]{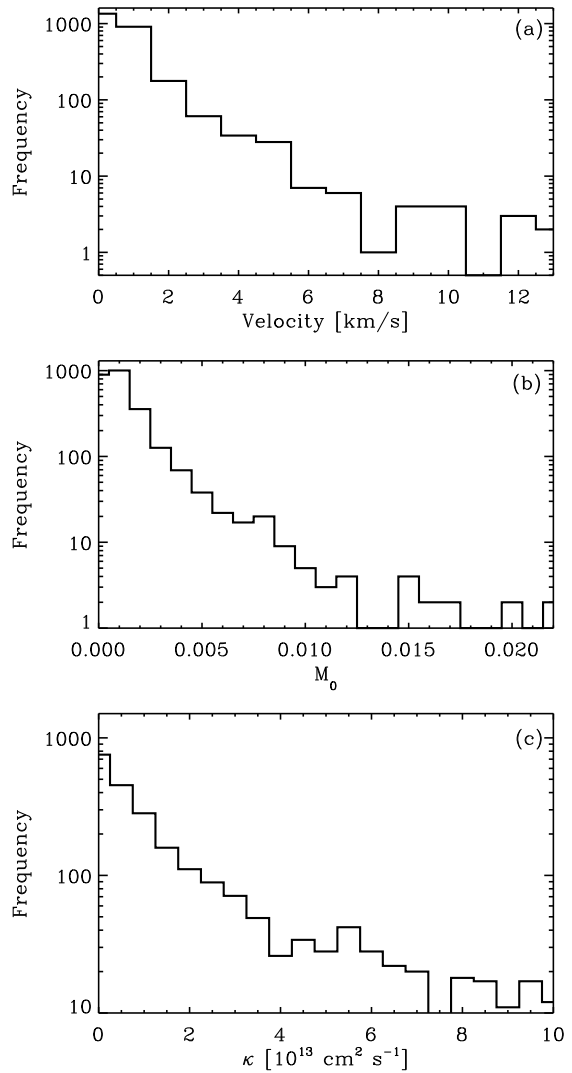} }
\vspace{-0.001\textwidth} 
\caption{Frequency distributions of boundary displacement velocities (a), magnetic reconnection rates (b), and diffusion coefficients (c).}
\label{fig4}
\end{figure}

We calculate the magnetic diffusion coefficient for each $5^{\circ}$ segment of each CH observation using Equation \ref{kappa} by substituting the boundary displacement distances in the segment ($\delta r$) and the time elapsed ($\delta t$) between consecutive observations. The resulting distribution of diffusion coefficients is shown in Figure~\ref{fig4}c, with values typically $\leq$3~$\times$~10$^{13}~$cm$^{2}~$s$^{-1}$. Using typical loop footpoint distances and characteristic times for reconnection, \cite{Fisk01} have determined the diffusion coefficients inside and outside of a CH to be 3.5~$\times$~10$^{13}$~cm$^{2}$~s$^{-1}$ and 1.6~$\times$~10$^{15}$~cm$^{2}$~s$^{-1}$, respectively. 

\section{Discussion and Conclusions}
\label{sec:conclusions} 

The relocation of open magnetic field lines through interchange reconnection is the most widely accepted method to explain some of the distinctive characteristics of CHs - namely their rigid rotation, the lack of loops with heights $>$10~Mm inside CHs, and the CH boundary relocations through interaction with QS loops. In this study CH boundary locations were determined using an intensity thresholding technique described in detail in \cite{Krista09}. This method determines the CH boundaries based on observations rather than allowing the observer to choose a subjective threshold. A number of CHs were tracked for several days to study the changes in the CH boundaries. The CH boundary tracking was carried out by temporally coupling the boundary contours, which allowed relative distances of boundary movements and hence velocities to be established. In our study the displacement velocities were found to be isotropic with unsigned values typically $\leq$2~km~s$^{-1}$, which imply reconnection rates of $\leq$3~$\times$~10$^{-3}$.

These boundary velocities are in good agreement with the isotropic nature and $\sim$1--2~km~s$^{-1}$ average velocity of random granular motions \citep{Markov91}. Hence, we suggest that the majority of the reconnection at CH boundaries is likely to be initiated through random granular motions. Velocities in excess of $\sim$2~km~s$^{-1}$ were occasionally observed during our analysis. Such surges in the boundary displacement could be explained by consecutive interchange reconnections in the case of favourable loop alignments, whereby a number of loops have opposite magnetic polarity to the CH flux in their nearest footpoint to the CH. In close proximity the CH open field lines could continuously reconnect with the loops and advance into the QS, before being halted by loops that face the CH with the same polarity in their nearest footpoint to the CH.

Interchange reconnection has been previously discussed by \cite{Fisk01} as a mechanism that allows the transfer of magnetic flux from one point to another (see Equation~\ref{eq_fisk}). We determine typical values of diffusion coefficient to be $\leq$3~$\times$~10$^{13}$~cm$^{2}$~s$^{-1}$ using Equation~\ref{kappa} with every individual CH boundary displacement measurement (i.e., from every 5 degree segment of each CH time pairing). Although the values derived in this study are comparable to those determined by \citeauthor{Fisk01} (3.5~$\times$~10$^{13}$~cm$^{2}$~s$^{-1}$ inside a CH), it is important to note that the results are achieved by different methods. Those presented here are based on CH boundary tracking using observations and reflect upon the diffusion processes happening at the CH boundary, while \citeauthor{Fisk01} provided estimates of the magnetic diffusion in CHs based on typical loop heights, foot-point distances and sizes of supergranular cells.

Previous studies found that the potential field source surface and magnetohydrodynamic models are successful in reconstructing the evolution of the global magnetic field through a sequence of steady-state solutions. Despite the quasi-steady and interchange reconnection models disagreeing about the existence of disconnected open fields, it is agreed that the highly dynamic short-term changes in the small-scale magnetic field may be explained by interchange reconnection between open and closed fields \citep{Antiochos07}. This could also be an alternative approach to explain how open field can be conglomerated (CH birth) or diffused (CH decay) on short-timescales. The local loop-emergence rate and the specific orientation of QS loops near the CH boundary may also influence the growth and decay of small CH candidates. Magnetic bipoles emerging within a CH or small loops admitted into the CH during interchange reconnection could accumulate over a CH lifetime and ultimately lead to the fragmentation and diffusion of the CH. Bipoles and open flux regions would have to be tracked in time to provide observational evidence for the processes governing the birth and the decay of CHs. This could also provide information on how small CHs turn into large, stable and long-lived CHs, and what processes initiate the decay of a long-lived CHs.

\acknowledgments
LDK is a Government of Ireland Scholar and has received funding from the Irish Research Council for Science, Engineering \& Technology (IRCSET) funded by the Irish National Development Plan. DSB is supported by the European Community (FP7) under a Marie Curie Intra-European Fellowship for Career Development. Images courtesy of {\it SOHO}.



{\it Facilities:} \facility{SOHO (EIT \& MDI)}

\clearpage



\end{document}